\documentclass[aps,twocolumn,superscriptaddress,floatfix,nofootinbib]{revtex4-1}
\usepackage{graphicx,amsmath,amssymb,verbatim,color}
\usepackage{booktabs}
\usepackage{comment}
\usepackage{soul}
\usepackage{setspace}
\usepackage[dvipsnames]{xcolor}
\usepackage{bm}
\usepackage[utf8]{inputenc}
\usepackage[colorlinks=true,citecolor=blue,linkcolor=blue,urlcolor=blue, backref=false,pdfborder={0 0 0}]{hyperref}
\usepackage{float}
\usepackage{multirow}
\usepackage{dcolumn}
\usepackage{aas_macros}
\usepackage[normalem]{ulem}
\usepackage{orcidlink}
\usepackage[capitalise]{cleveref}

\begin{document}

\preprint{APS/123-QED}

\title{A frequentist view on the two-body decaying dark matter model}

\author{Thomas Montandon}
\affiliation{Laboratoire Univers \& Particules de Montpellier, CNRS \& Université de Montpellier (UMR-5299), 34095 Montpellier, France}
 \email{thomas.montandon@umontpellier.fr}

\author{Elsa M. Teixeira}
\affiliation{Laboratoire Univers \& Particules de Montpellier, CNRS \& Université de Montpellier (UMR-5299), 34095 Montpellier, France}

\author{Adèle Poudou}
\affiliation{Laboratoire Univers \& Particules de Montpellier, CNRS \& Université de Montpellier (UMR-5299), 34095 Montpellier, France}

\author{Vivian Poulin}
\affiliation{Laboratoire Univers \& Particules de Montpellier, CNRS \& Université de Montpellier (UMR-5299), 34095 Montpellier, France}

\begin{abstract}
Decaying dark matter (DDM) has emerged as an interesting framework to extend the $\Lambda$-cold-dark-matter ($\Lambda$CDM) model, as many particle physics models predict that dark matter may not be stable over cosmic time and can impact structure formation. In particular, a model in which DM decays at a rate $\Gamma$ and imprints a velocity kick $v$ onto its decay products leads to a low amplitude of fluctuations, as quantified by the parameter $S_8$, in better agreement with that measured by some past weak lensing surveys. Bayesian analyses have provided mixed conclusions regarding its viability, with a reconstructed clustering amplitude only slightly below the standard $\Lambda$CDM value. In this paper, we contrast previous results with a frequentist analysis of Planck and SDSS BAO data. We find that the $68\%$ confidence level region corresponds to a decay half-life of $6.93^{+7.88}_{-2.85}\, \mathrm{Gyr}$ and a velocity kick of $1250^{+1450}_{-1000}$~km/s. These $1\sigma$ constraints strongly differ from their Bayesian counterparts, indicating the presence of volume effect in the Bayesian analysis. Moreover, we find that under the DDM model, the frequentist analysis predicts lower values of $S_8$, in agreement with those found by \texttt{KiDS-1000} and \texttt{DES-Y3} at $\sim 1.5\sigma$. We further show that previously derived \texttt{KiDS-1000} constraints that appeared to exclude the best-fit model from {\it Planck} data were driven by priors on the primordial amplitude $A_s$ and spectral index $n_s$. When those are removed from the analysis, \texttt{KiDS-1000} constraints on the DDM parameters are fully relaxed. It is only when applying \textit{Planck}-informed priors on $A_s$ and $n_s$ to the \texttt{KiDS-1000} analysis that one can constrain the model. 
We further highlight that in the absence of such priors, the region of scales best-measured by \texttt{KiDS-1000} does not exactly match the $S_8$ kernel (centred around $k\sim 0.1\, h/$Mpc), but rather a slightly smaller range of scales centred around $k\sim 0.3\, h/$Mpc. One must thus be careful in applying $S_8$ constraints to a model instead of the full data likelihood.

\end{abstract}

\maketitle

\section{Introduction} 
Over the past decades, the $\Lambda$CDM (``Lambda'' Cold Dark Matter) cosmological model has emerged as the standard paradigm for describing the Universe, providing an excellent fit to a wide range of observations spanning both early and late cosmic times. From the precise measurements of the Cosmic Microwave Background (CMB) and Big Bang Nucleosynthesis (BBN) to large-scale structure surveys, baryon acoustic oscillations (BAO), and supernovae type Ia (SNIa), the $\Lambda$CDM framework has proven remarkably successful. Nonetheless, despite its empirical success, the fundamental nature of its dominant constituents -- cold dark matter (CDM) and dark energy ($\Lambda$) -- remains unknown, and tensions have emerged as the precision of cosmological data has improved.

Among the most prominent of these discrepancies is the so-called $S_8$ tension, referring to the $2-3\sigma$ disagreement between the amplitude of matter fluctuations inferred from high-redshift CMB data and direct measurements at low redshift from weak lensing (WL) surveys. The parameter $S_8=\sigma_8 \sqrt{\Omega_{\rm m}/0.3}$, where $\sigma_8$ quantifies the matter fluctuation amplitude on $8\, h^{-1}$Mpc scales and $\Omega_{\rm m}$ the total matter density, encapsulates this tension under $\Lambda$CDM. For instance, the {\it Planck} satellite has measured $S_8 = 0.825 \pm 0.011$ \cite{Aghanim:2018eyx}, while weak lensing surveys like the Kilo-Degree Survey (KiDS) \cite{Kuijken:2019gsa, Giblin:2020quj}, the Dark Energy Survey (DES) \cite{DES:2021bvc}, and the Hyper Supreme-Cam (HSC) \cite{Hamana:2019etx} measure consistently lower values, such as $S_8=0.748^{+0.021}_{-0.025}$ in \texttt{KiDS-1000} \cite{Busch:2022pcx}. However, more recent analyses such as \texttt{DES-Y3} \cite{DES:2021wwk} and the \texttt{KiDS-Legacy} results \cite{Stolzner:2025htz} strongly mitigate the existence of a $S_8$ tension. The large constraining power of weak lensing will be vastly used by the Vera Rubin observatory \cite{LSST:2008ijt}, \textit{Euclid} \cite{Euclid:2024yrr}, Nancy Grace Roman Space Telescope \cite{Spergel:2015sza} and the Chinese Space Station Telescope \cite{Gong:2019yxt}. Hence, we should expect definitive answers regarding small-scale physics in the next few years. However, whether or not there is an $S_8$ tension, constraining physically motivated models remains crucial, given that the nature of CDM is still entirely unknown.

While an $S_8$ tension might also emerge from primordial non-Gaussianities \cite{Stahl:2024stz}, or small-scale physics such as baryonic feedback or nonlinear structure formation \cite{Tan:2022wob, Amon:2022azi, Arico:2023ocu}, it could also point to the need for new physics beyond the standard model. Among the many proposed extensions to $\Lambda$CDM, a particularly well-motivated class involves modifying the properties of dark matter itself, see \textit{e.g.} Refs.~\cite{Schneider:2019xpf, Heimersheim:2020aoc, Joseph:2022jsf, Poulin:2022sgp, Ferlito:2022mok} for recent models. In this context, decaying dark matter (DDM) scenarios \cite{Enqvist:2015ara, Enqvist:2019tsa, Murgia:2017lwo, Abellan:2021bpx} have received growing attention \cite{DES:2020mpv, Choi:2021uhy, Tanimura:2023bkh, Bucko:2022kss, Bucko:2023eix}, as they can naturally predict a suppression of power on small scales ($k \sim 0.1 - 1\, h/$Mpc), thereby alleviating the $S_8$ tension. 

Many particle physics-motivated scenarios of dark matter \cite{Hambye:2010zb, Abazajian:2012ys, Drewes:2016upu, Doroshkevich:1984gw, Doroshkevich:1989bf, Khlopov:1995pa, Berezinsky:1991sp, Covi:1999ty, Kim:2001sh, Chou:2003wx, Feng:2003uy, Ghosh:2020ipv, Dutta:2022wuc, Fuss:2024dam} question its stability on cosmological timescales and can effectively be described by a two-body DDM model. While the DDM framework has attracted attention for its potential to ease the $S_8$ tension, disfavoring it -- as suggested by recent weak lensing measurements from \texttt{DES-Y3} \cite{DES:2021wwk} and \texttt{KiDS-Legacy} \cite{Stolzner:2025htz} -- would have broader implications, potentially ruling out a wide class of unstable dark matter models. 

The most simple class of DDM considers the decay of CDM into dark radiation, for which strong constraints can be obtained with CMB data \cite{Nygaard:2020sow, Holm:2022kkd, Simon:2022ftd, Bucko:2023eix}. Another kind of interesting variant is the two-body decaying dark matter model \cite{Wang:2012eka}, in which a fraction of the CDM decays into a massive daughter particle and a massless component (\textit{e.g.} dark radiation). The massive product inherits a velocity kick from energy-momentum conservation, leading to a scale-dependent suppression of the matter power spectrum due to its free-streaming. This mechanism can alter structure formation like warm dark matter, but with a redshift and scale dependence that can leave the early time unchanged compared to CDM.

Previous studies have investigated the phenomenology of the DDM model using linear perturbation theory \cite{Wang:2012eka, Abellan:2021bpx} or simulations \cite{Peter:2010sz, Cheng:2015dga}. Constraints have been performed using the {\it Planck} CMB data, BAO and supernova \cite{Abellan:2021bpx}, BOSS full shape \cite{Simon:2022ftd},
Milky Way satellite counts \cite{DES:2022doi}, and Lyman-$\alpha$ forest observations \cite{Wang:2013rha, Fuss:2022zyt}. More recently, Ref.~\cite{Bucko:2023eix} used the weak lensing survey \texttt{KiDS-1000} to derive the strongest constraints obtained. 
As far as the two-body decay is concerned -- see Ref.~\cite{Holm:2022kkd} for a profile likelihood analysis of the one-body DDM model decaying into radiation -- all these studies have used the standard Bayesian framework which is known to be affected by prior volume effects \cite{Nygaard:2023cus}. 

In this work, we further explore the DDM scenario in light of the latest cosmological observations. We perform an independent analysis of \texttt{KiDS-1000} cosmic shear data using a nonlinear matter power spectrum emulator developed in Ref.~\cite{Bucko:2023eix}, and reanalyse the {\it Planck} 2018 CMB data, adding also BAO and supernovae, to extract updated constraints using a frequentist analysis to get rid of any prior volume effects. We also combine both datasets using {\it Planck} informed priors to assess the potential of this scenario to reconcile CMB and WL observations. 

This paper is structured as follows: \cref{sec:model_method} introduces the DDM model, the dataset and the methodology used. In \cref{sec:bayesVSfreq}, we present the standard Bayesian results and the {\it Planck} likelihood profiles analysis. \cref{sec:KiDS} discusses the constraints of \texttt{KiDS-1000} and proposes a consistent \texttt{KiDS-1000} and {\it Planck} joint analysis. The conclusions are presented in \cref{sec:concl}.

\section{Model and Methodology}\label{sec:model_method}
The standard cosmological model, flat $\Lambda$CDM, is characterised by a set of fundamental parameters describing the evolution, composition and initial conditions of the Universe. These include the Hubble constant $H_0$, the cold dark matter density parameter $\Omega_c$, the baryon energy density $\Omega_b$, and the photon energy density, which is now well constrained through the monopole temperature of the CMB, measured to be $T_0 = 2.7255 \pm 0.0006\,$K \cite{Fixsen:2009ug}. Motivated by single-field inflation models and observational constraints performed by {\it Planck} \cite{Planck:2019kim, Planck:2018jri}, the primordial initial conditions are assumed to be Gaussian and adiabatic perturbations, with a power spectrum following a power law with amplitude $A_s$ and spectral index $n_s-1$. 

We now focus on the scenario where CDM particles are unstable and can decay into a lighter particle and a massless relativistic particle. 

\subsection{Two-Body Decaying Dark Matter}\label{sub:DDM}

In this work, we focus on the DDM model in which a CDM particle of mass $M_{\rm CDM}$ decays into two lighter particles: a massive ``warm'' daughter particle of mass $M_{\rm WDM}$ and a massless relativistic particle. The decay is characterised by the decay rate $\Gamma$, which is inversely proportional to the particle's lifetime. The warm daughter particle inherits a velocity kick $v$. Given by the kinematic relation of the decay, the fraction $\epsilon$ of rest-mass energy of the mother particle converted into kinetic energy reads:
\begin{equation}
\epsilon = \frac{1}{2} \left(1-\frac{M_{\rm WDM}^2}{M^2_{\rm CDM}} \right)\,.
\end{equation}
In the non-relativistic limit, the velocity kick is given by $v = \epsilon c$, where $c$ is the speed of light.
A third parameter, $f$, represents the fraction of unstable CDM that decays into WDM. In this analysis, we focus on the case $f=1$, where all the CDM decays into WDM. A thorough account of the phenomenology of this model and the corresponding set of dynamical equations are presented in Refs.~\cite{Blackadder:2014wpa, Abellan:2021bpx}.

Note that the $\Lambda$CDM limit is recovered when the decay lifetime greatly exceeds the age of the Universe ($\Gamma^{-1}~\to~\infty$) or when the momentum transfer is negligible ($\epsilon\to0$), leaving the daughter particle effectively cold. The limit $\epsilon \to 0.5$ ($M_{\rm WDM} \to 0$) corresponds to a one-body decay with the daughter particle acting effectively as dark radiation.  

The momentum transfer to the daughter particle generates a time- and scale-dependent free-streaming scale. 
Below this scale, density perturbations are suppressed because the pressure of the warm daughter particle counteracts gravitational collapse. The time dependence of this effect is governed by the dark matter lifetime $\sim \Gamma^{-1}$, such that only perturbation modes below the free-streaming scale are significantly affected after times $t > \Gamma^{-1}$. 
Consequently, perturbations evolve similarly to those of standard CDM for times $t < \Gamma^{-1}$ and on scales larger than the free-streaming scale. 
This allows the model to reproduce the CMB temperature, polarisation and lensing anisotropies with a CDM-like behaviour while still reducing small-scale power in the late-time matter power spectrum due to the suppression of growth at small scales. 
The depth and the scale of the matter power suppression are controlled by $\Gamma$ and $v$, respectively.

The growth of cosmic structures is often quantified by the variance of matter fluctuations smoothed over a scale $r$, defined as:
\begin{equation}\label{eq:sigma8_def}
\sigma_r^2 = \int_0^\infty \frac{k^2 dk}{2\pi^2} P_m(k) W^2(r, k)\,,
\end{equation}
where $W$ is the Fourier transform of a top-hat filter, $k$ is the corresponding wavenumber in Fourier space and $P_m(k)$ is the matter power spectrum:
\begin{equation}\label{eq:Pk}
\left<\delta_m(\boldsymbol k_1) \delta_m(\boldsymbol k_2)\right> = (2\pi)^3 \delta^{\rm D} (\boldsymbol k_1 + \boldsymbol k_2 ) P_m(k_1)\,,
\end{equation}
where $\delta_m$ is the total matter density contrast, and $\delta^{\rm D}$ is the Dirac-delta function. The characteristic scale of cosmic-structure formation is usually taken to be $r \sim 8 h^{-1}$~Mpc, in which case we refer to $\sigma_8$ in \cref{eq:sigma8_def}. The key signature of the warm dark matter species is to predict a smaller value of $\sigma_8$ than CDM. However, weak lensing measurements are sensitive to both the amplitude of matter fluctuations, $\sigma_8$, and the total matter density, $\Omega_{\rm m}$, leading to a degeneracy between these parameters. This motivates the use of the derived parameter
\begin{equation}\label{eq:S8}
S_8 = \sigma_8 \sqrt{\frac{\Omega_{\rm m}}{0.3}}\,.
\end{equation}

The linear evolution of the DDM model has been implemented in the Einstein-Boltzmann solver \texttt{CLASS} \cite{Lesgourgues:2011re, Blas:2011rf} and made publicly available \cite{Abellan:2021bpx}. This allows efficient numerical computation of the linear power spectrum and the CMB angular power spectra $C_\ell$:
\begin{equation}
\left< a^X_{\ell_1 m_1} a^Y_{\ell_2 m_2}\right> =  \delta^{\rm K}_{\ell_1\ell_2} \delta^{\rm K}_{m_1m_2} C^{XY}_{\ell_1}\,,
\end{equation}
where $\delta^{\rm K}$ is the Kronecker-delta, and $X,Y$ denote temperature, polarisation, or lensing fields.

For late-time probes, such as the case of galaxy weak lensing, which we will consider, nonlinear corrections are required. The nonlinear matter power spectrum is often computed using \texttt{Halofit} \cite{Takahashi:2012em, Smith:2002dz}, calibrated on $\Lambda$CDM simulations. DDM N-body simulations have been performed in Ref.~\cite{Cheng:2015dga}. Here, the latest nonlinear power spectrum emulation of Ref.~\cite{Bucko:2023eix} is employed. Their emulator, \texttt{DMemu} \cite{Bucko:2023eix}, provides the nonlinear suppression factor:
\begin{equation}\label{eq:pk_nl}
P_{\rm NL}^{\Lambda \rm DDM} (k, z) = P_{\rm NL}^{\Lambda \rm CDM}(k, z)  \times  S^{\Gamma, v}(k, z)\,.
\end{equation}

A significant source of potential degeneracies is baryonic effects, which also affect small-scale power, necessitating careful modelling \cite{Giri:2021qin}. 
To include them, we use the same methodology as in Ref.~\cite{Kilo-DegreeSurvey:2023gfr}, which consists of marginalising over the baryonic feedback parameter using \texttt{HMcode2020} \cite{Mead:2020vgs}. 

The cosmic shear angular power spectrum, computed using the Limber approximation \cite{KiDS:2020suj}, is given by:
\begin{equation}
C^{(ij)}_{\ell} = \int_{0}^{\chi_{\rm h}} \frac{d\chi}{\chi^2} W_{\Phi}^{(i)}(\chi) W_{\Phi}^{(j)}(\chi) P_{\rm{NL}}\left(\frac{\ell + 1/2}{\chi}, z(\chi)\right)\,,
\end{equation}
where $\chi_{\rm h}$ is the comoving horizon distance, and $W_{\Phi}^{(i)}$ is the lensing kernel of the tomographic bin $(i)$. For details on intrinsic alignment modelling, see Ref.~\cite{Joachimi:2020abi}.

\subsection{Datasets and Methodology}

In this work, we use two datasets labelled ``{\it Planck}'' and ``\texttt{KiDS-1000}''.
\begin{itemize}
    \item {\it Planck} refers to the standard {\it Planck} CMB data compiled in the Plik 2018 likelihood \cite{Planck:2019nip}, including high- and low-$\ell$ temperature, polarisation, and cross-correlations, as well as lensing \cite{Planck:2018lbu}. In addition, we include the BAO measurements from the low-redshift surveys 6dFGS \cite{Beutler:2011hx} and SDSS DR7 \cite{Ross:2014qpa}, as well as the full BOSS DR12 \cite{BOSS:2016wmc}. Finally, we include the Pantheon-Plus compilation of Type Ia supernovae distance moduli measurements \cite{Brout:2022vxf}.
    \item the Kilo-Degree Survey \texttt{KiDS-1000} \cite{KiDS:2020suj}, also labeled as ``KiDS'', is a galaxy catalog providing information about $\sim 20$ million galaxies divided into $5$ redshift bins covering a range of $z \in [0.1, 1.2]$. The catalogue is analysed by the European Southern Observatory with the Very Large Telescope (VLT). We use the likelihood \texttt{COSEBIs} \cite{Schneider:2010pm, Asgari:2012ir, Asgari:2016xuw}.\footnote{Short before submission of this paper, the last data release was performed under the name \texttt{KiDS-Legacy} \cite{Stolzner:2025htz}.}
\end{itemize}

To analyse these two datasets within the theoretical framework of \cref{sub:DDM}, we first perform a Bayesian analysis using the standard Markov Chain Monte Carlo (MCMC) method \cite{Metropolis:1953am, Hastings:1970aa}. For the {\it Planck} analysis, we use \href{https://github.com/brinckmann/montepython_public}{\texttt{MontePython}} \cite{Brinckmann:2018cvx, Audren:2012wb}. For practical reasons, we use \href{https://cosmosis.readthedocs.io/en/latest/index.html}{\texttt{CosmoSIS}} \cite{Zuntz:2014csq}, which directly includes the \texttt{COSEBIs} likelihood, as well as more recent likelihoods such as \texttt{DES-Y3} \cite{DES:2021wwk}, and their joint analysis \cite{Kilo-DegreeSurvey:2023gfr} which will be used in future works. We vary all 6-$\Lambda$CDM parameters within broad flat priors. This is particularly important for $A_s$ and $n_s$, as we show later on. 
The prior ranges for the DDM parameters used in our analysis are chosen to match those adopted in previous studies that derived $1$- and $2\sigma$ constraints on the DDM model using {\it Planck} data (e.g., \cite{Abellan:2020pmw}). In these works, log-flat priors were employed to cover a wide dynamic range while remaining agnostic about the characteristic scale of deviations from standard CDM. Following this approach, we adopt
\begin{align}
\log_{10} \Gamma \times \mathrm{Gyr} &\in [-4, 1]\,,\nonumber\\
\log_{10} v \times \mathrm{s/km} &\in [1.5, 5]\,.
\end{align}
This corresponds to decay half-lives in the range $[0.1, 10^4]$~Gyr -- broad enough to include scenarios where decay is negligible over the age of the Universe -- and to velocity kicks from $30$~km/s (comparable to the escape velocity of Saturn) up to $10^5$~km/s ($\sim c/3$).
Since current data constrain only a combination of $\Gamma$ and $v$, and given that training the matter power spectrum emulator require to run DDM and CDM simulations increasing the computational cost, a more restricted prior volume was chosen by Ref.\cite{Bucko:2022kss} to train the emulator $S^{\Gamma, v}$, see \cref{eq:pk_nl}. Hence, for \texttt{KiDS-1000}, we adopt the same priors 
\begin{align}
\log_{10} \Gamma \times \mathrm{Gyr} &\in [-4, -1.13]\,,\nonumber\\
\log_{10} v \times \mathrm{s/km} &\in [1.5, 3.7]\,,
\end{align}
which still captures the $2\sigma$ constraints reported in Refs.\cite{Abellan:2021bpx, Simon:2022ftd, DES:2022doi, Wang:2013rha, Fuss:2022zyt}. This corresponds to half-lives up to $\sim 9.4$~Gyr and velocity kicks up to $5000$~km/s.

Bayesian analyses can suffer from strong prior dependence. Even though the intrinsic interpretation of Bayesian and frequentist approaches differ and cannot strictly be compared, frequentist analyses provide prior-independent constraints that help disentangle the impact of the priors from that of the likelihood in Bayesian inferences.

To that end, frequentist analyses employ the ``profile likelihood'' method. Instead of marginalising over nuisance parameters as in Bayesian MCMC, the profile likelihood for a parameter set $\boldsymbol \alpha$ is defined as:
\begin{equation}\label{eq:def_profile}
\mathcal{L}_p(\boldsymbol \alpha) = \max_{\boldsymbol \theta} \mathcal{L}(\boldsymbol \alpha, \boldsymbol \theta)\,,
\end{equation}
where $\mathcal{L}(\boldsymbol \alpha, \boldsymbol\theta)$ is the full likelihood function, $\boldsymbol \theta$ represents all parameters to be minimised and $\boldsymbol \alpha$ the subset of parameter to study. In our specific case, the parameters of interest $\boldsymbol{\alpha}$ are taken to be either $\Gamma$, $v$ independently, or both jointly $\Gamma$ and $v$, while $\boldsymbol \theta$ includes all remaining cosmological and nuisance parameters over which the likelihood is maximised. For Gaussian likelihoods, we have $-2\log \mathcal L_p(\boldsymbol \alpha) = \chi_{\boldsymbol  \alpha}^2$ which allows us to determine confidence intervals from $\Delta \chi^2 (\boldsymbol \alpha) = \chi^2({\boldsymbol \alpha}) - \chi^2_{\rm DDM}$ where $\chi^2_{\rm DDM}$ here represents the global maximum likelihood of the DDM model, hence taking $\boldsymbol \alpha$ also as free parameter. To compare the improvement of a model with respect to a reference model -- here $\Lambda$CDM -- we compare their global minima $\Delta \chi_{\rm DDM}^2 = \chi^2_{\rm DDM} - \chi^2_{\rm CDM}$. 
We minimise using a simulated annealing method described in Ref.~\cite{Hannestad:2000wx}.

\section{Bayesian vs Frequentist Analyses} \label{sec:bayesVSfreq}
\subsection{Bayesian constraints from {\it Planck} and \texttt{KiDS-1000}} \label{sec:bayes}

\begin{figure}
    \centering
    \includegraphics[scale=0.6]{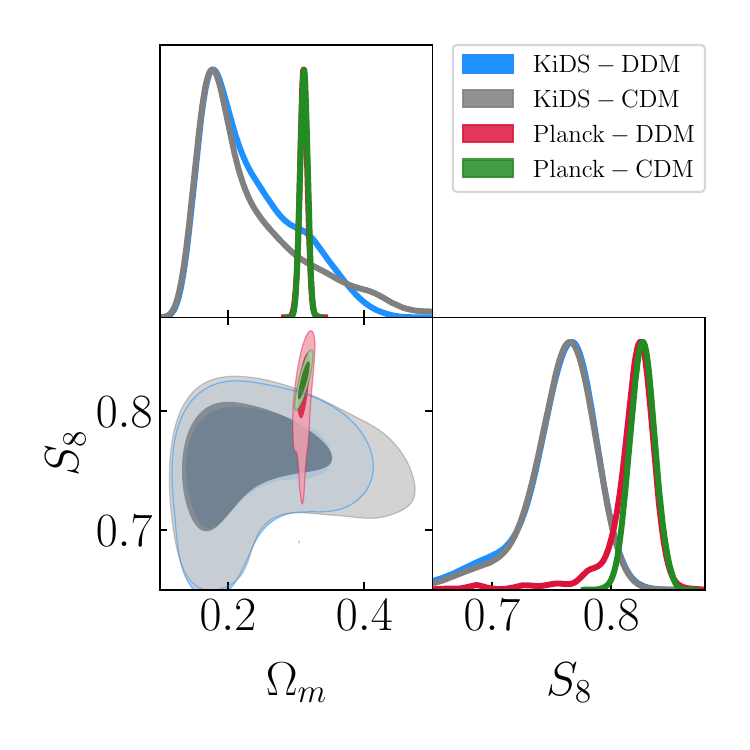}
    \caption{Constraints on the cosmological parameters $S_8$ and $\Omega_{\rm m}$ from \texttt{KiDS-1000} and {\it Planck} data for both the Decaying Dark Matter (DDM) and Cold Dark Matter (CDM) models. The contours show the 1$\sigma$ and 2$\sigma$ confidence regions. This comparison highlights the impact of the DDM model on bringing large-scale structure constraints into closer agreement with \textit{Planck}. 
    }
    \label{fig:bayesian}
\end{figure}

In \cref{fig:bayesian}, we reproduce the standard Bayesian analysis for {$\Lambda$}DDM and {$\Lambda$}CDM using the two aforementioned data sets. We show the 2D marginalised posteriors for the parameters $\Omega_{\rm m}$ and $S_8$ in the two models: DDM in blue and red, for \texttt{KiDS-1000} and \textit{Planck}, respectively, and grey and green for the standard $\Lambda$CDM model for \texttt{KiDS-1000} and {\it Planck}, respectively. In agreement with what has been discussed in past literature, \textit{e.g.} Ref.~\cite{KiDS:2020suj, Bucko:2023eix}, we can see how \texttt{KiDS-1000}, as well as {\it Planck}, independently provide precise measurements of $S_8$.  
Famously, the so-called $S_8$ tension appears as a $\sim 2\sigma$ discrepancy between {\it Planck} and \texttt{KiDS-1000} for the $\Lambda$CDM model. 
In that regard, the DDM model helps decrease the tension by extending the $2\sigma$ {\it Planck} contour toward the low-$S_8$ \texttt{KiDS-1000} contours. This apparently small improvement should be taken with caution because the DDM parameter space is not well constrained and can, therefore, be subject to a significant volume effect that alters the reconstructed posteriors in a Bayesian analysis. Moreover, as mentioned in the introduction, more recent data such as the \texttt{DES-Y3} \cite{DES:2021wwk} and the \texttt{KiDS-Legacy} results \cite{Stolzner:2025htz} do not show any tension between these two datasets. What we see here is, therefore, likely due to systematic effects in the \texttt{KiDS-1000} analysis pipeline (see \cite{Wright:2025xka} for details).

\subsection{Planck Profile Likelihood}\label{sec:Planck_profile}

We now compute the 1- and 2-dimensional profile likelihood of the two additional parameters of the DDM model $\Gamma$ and $v$ for the {\it Planck} data set and discuss their consistency with the \texttt{KiDS-1000} measurements of $S_8$. 
We focus in the main text on the case of {\it Planck} data because we have found that the \texttt{KiDS-1000} profile is strongly affected by the boundary set to one of the nuisance parameters, making difficult the interpretation of the frequentist analysis of \texttt{KiDS-1000}. 
Concretely, the galaxy intrinsic alignment model parameters (\textit{i.e.} the intrinsic alignment redshift dependence $\alpha_1$ in \texttt{CosmoSIS} and $\eta_1$ in Ref.~\cite{Kilo-DegreeSurvey:2023gfr}) seems inconsistent between \texttt{KiDS-1000} or \texttt{DES-Y3} compared to direct observation, see footnote $36$ in Ref.~\cite{Kilo-DegreeSurvey:2023gfr}. A flat prior is adopted on this parameter, making it reach the boundaries during the minimisation procedure, potentially spoiling the results, which at the same time reach the upper bound of the prior on $v$. We present this analysis in \cref{app:kids_pl} and focus here on the case of {\it Planck}.

We first run a global minimisation and find a global improvement of the DDM fit compared to $\Lambda$CDM of 
\begin{equation}
\Delta \chi^2_{\rm Planck, DDM} = -2.66\,,
\end{equation}
corresponding to values 
\begin{equation}
    \Gamma = 0.10\,\mathrm{Gyr^{-1}},\,~~ v=1250\, \mathrm{km/s},\,~~ S_8 = 0.70\,.
    \label{eq:values}
\end{equation}
Interestingly, this value of $S_8$ lies at the edge of the Bayesian posterior reconstructed from the analysis of {\it Planck} data (from \cref{fig:bayesian} we obtain $S_8 = 0.816 \pm 0.028$), indicating the presence of a significant volume effect as anticipated. In fact, it is even below the interval measured by \texttt{KiDS-1000}. Computing the profile likelihood will thus allow us to derive prior-independent confidence intervals and robustly establish the extent to which the DDM model can reduce $S_8$ and fit both {\it Planck} and \texttt{KiDS-1000} data.

\begin{figure}
    \centering
    \includegraphics[scale=0.45]{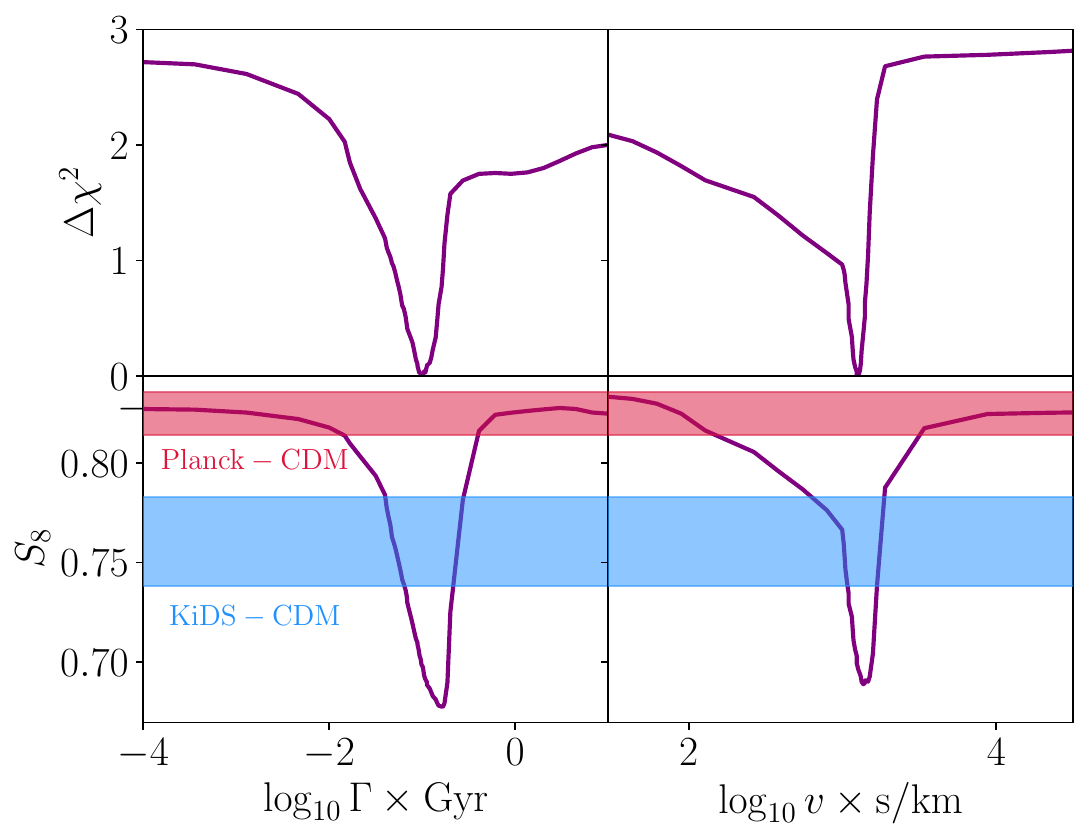}
    \caption{Planck profile likelihood as a function of $\log_{10} \Gamma$ on the left column and $\log_{10} v$ on the right column. The top panels show $\Delta \chi^2$ while the bottom panels show the $S_8$ associated with the best-fit. The red band indicates the {\it Planck} 2018 constraints on $S_8$ from Ref.~\cite{Aghanim:2018eyx}, while the blue bands indicate the \texttt{KiDS-1000} constraints from Ref.~\cite{KiDS:2020suj}. }
    \label{fig:profile_1d}
\end{figure}
\begin{figure}
    \centering
    \includegraphics[scale=0.7]{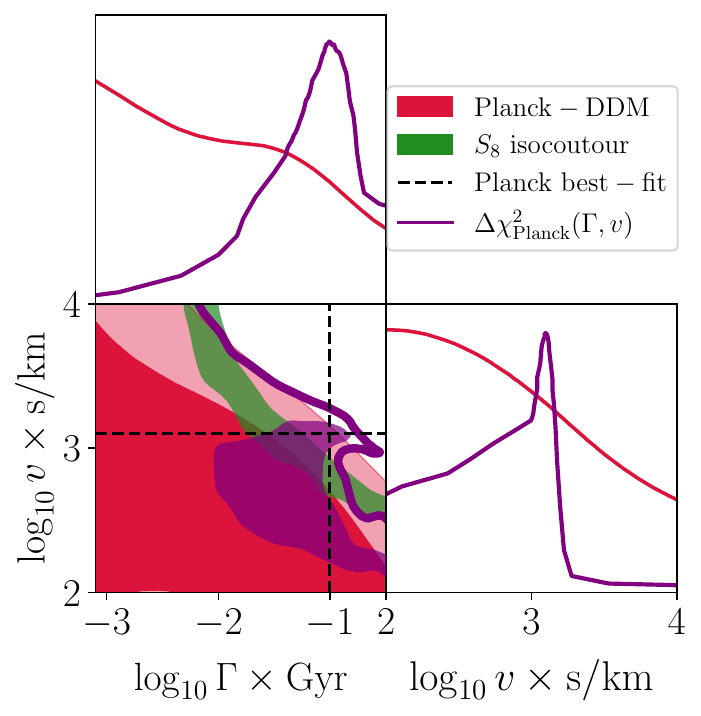}
    \caption{Constraints on the two additional parameters of the DDM model for {\it Planck}. The red contours show the result of the Bayesian analysis. The violet line shows the profile likelihood 2$\sigma$ region, while the filled violet contour represents the $1\sigma$ confidence level. The green band represent the $S_8$ isocontour for the upper and lower bounds of \texttt{KiDS-1000} \cite{Heymans:2020gsg}.}
    \label{fig:profile}
\end{figure}
\begin{figure*}[t]
    \centering
    \includegraphics[scale=0.6]{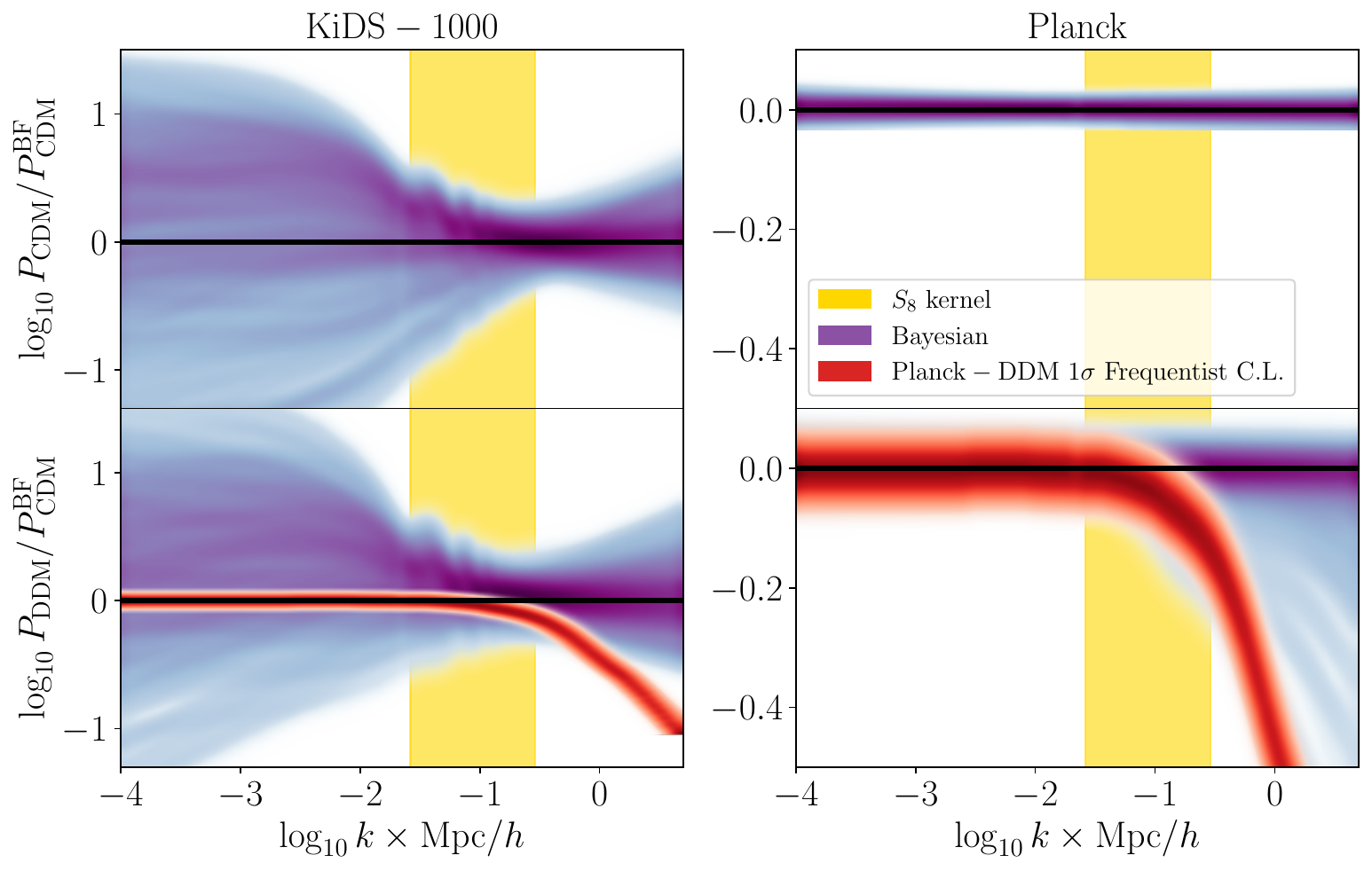}
    \caption{Bayesian posterior density of the power spectrum (in blue) and approximate frequentist $1\sigma$ confidence interval (in red; see main text for details), both normalized to the Planck CDM best-fit. In the left (right) panels for \texttt{KiDS-1000} ({\it Planck}), and in the first (second) row for the CDM (DDM) model. 
    The yellow band represents the main support of the $S_8$ kernel defined as $W(8 h/\mathrm{Mpc}, k) > 0.1$, see \cref{eq:sigma8_def}.}
    \label{fig:Pk_density}
\end{figure*}
\subsubsection{1-dimensional profile}
In \cref{fig:profile_1d}, we show the profile likelihood as a function of $\Gamma$ and $v$ on the first and second columns, respectively, by using \cref{eq:def_profile}. The top and bottom panels show respectively the $\Delta \chi^2$ and the derived parameter $S_8$ computed at the maximum likelihood $\boldsymbol \theta_{\rm max}$. We show the reconstructed value of $S_8$ in the \texttt{KiDS-1000} analysis as a blue band and the {\it Planck} $\Lambda$CDM $S_8$ prediction as a red band in the bottom panels of \cref{fig:profile_1d}. Confidence intervals can be extracted from the profile using the Neyman construction~\cite{Neyman:1937uhy}, which for 1D profile amounts to computing $\Delta\chi^2<1~(4)$ at 68\% (95\%) C.L. This procedure yields
\begin{align}
    \Gamma &= 0.10^{+0.17}_{-0.05}\,\mathrm{Gyr}^{-1} \,,\nonumber\\
    t^{\mathrm{DDM}}_{1/2} & = 6.93^{+7.88}_{-2.85}\, \mathrm{Gyr} \,,\nonumber\\
    N^{\mathrm{DDM}}_0 / N_{\rm ini} & = 0.25^{+0.27}_{-0.16} \,,\nonumber\\
    v   &= 1250^{+1450}_{-1000}\, \mathrm{km/s} \,,
\end{align}
where we have defined $t^{\mathrm{DDM}}_{1/2}$ to be the half-life of DDM and $N^{\mathrm{DDM}}_0$ and ${N_{\rm ini}}$ are respectively the present and initial number of DDM particles. Interestingly, rather than exclusion, the profile likelihood suggests that a specific region of values of $\Gamma$ and $v$ is weakly favoured over the $\Lambda$CDM model, although the $\Lambda$CDM limit is obtained at $1.6\sigma$. Hence, our best-fit and $68\%$ credible interval give us a half-life of $\sim 7$~Gyr, which means that $\sim 75\%$ of the initial DDM particles would have already decayed today. We anticipate that including information from the full shape galaxy power spectrum would alter those constraints, see \textit{e.g.} Ref.~\cite{Simon:2022ftd}, and we leave a dedicated combined analysis for future work.

Since low values of $\Gamma$ indicate a large lifetime compared to the age of the Universe while low $v$ suggests that the decay products remain cold, we trivially recover in these two limits the CDM case and its $S_8$ $\Lambda$CDM value. For too large values of $\Gamma $ or $v$, corresponding to either quick decays or hot decay products, the $\Lambda$CDM limit is also recovered: the algorithm simply adjusts whichever decay parameter is allowed to freely vary to its CDM limit to prevent a degradation of the fit. Consequently, one recovers again the $\Lambda$CDM value of $S_8$.
Importantly, one can see that the \texttt{KiDS-1000} $S_8$ band corresponds to consistently better $\chi^2$ values than the {\it Planck} $\Lambda$CDM $S_8$ band. This suggests that the DDM model can indeed accommodate the low $S_8$ value measured by \texttt{KiDS-1000} despite the apparent Bayesian constraints.

\subsubsection{2-dimensional profile}

We show in \cref{fig:profile} the comparison between Bayesian and frequentist intervals reconstructed from the analysis of {\it Planck} data in the 2-dimensional $\{\log_{10}\Gamma,\log_{10} v\}$ plane, as well as in the 1-dimensional case for each parameter. In the frequentist case, shown in purple, we display the 2-dimensional confidence intervals derived with the Neyman construction at  $1\sigma$ ($\Delta \chi^2 = 2.3$) and $2\sigma$ ($\Delta \chi^2 = 6.17$) and the $1$-dimensional profile of \cref{fig:profile_1d} normalised to unity. The marginalised Bayesian posteriors reconstructed from {\it Planck} are instead shown in red and are similar to those presented in Refs.~\cite{Bucko:2023eix, Abellan:2020pmw}. 
In green, we show the isocontours of $S_8$ for the lower and upper limits of the \texttt{KiDS-1000} measurements $S^{\rm KiDS}_8 = 0.759^{+0.024}_{-0.021}$, similar to the blue band of \cref{fig:profile_1d}. The global minimum is highlighted with the black dashed line, and we recall that it is associated with a low value of $S_8=0.70$, according to \cref{eq:values}.

First and foremost, one can see that the 1$\sigma$ confidence intervals differ from their Bayesian counterparts, confirming the presence of a volume effect. The frequentist interval reveals a closed contour at $1\sigma$, indicating a slight preference for a specific DDM parameter region within {\it Planck} data. However, the improvement in the fit is too small to overcome the large prior volume, disfavoring the DDM model in the Bayesian analysis. This is similar to what was found in Ref.~\cite{Holm:2022kkd} in a decaying dark matter model with massless decay products.

Second, it is striking that the $S_8$ values reconstructed from the \texttt{KiDS-1000} analysis (green band) intersect the $1\sigma$ confidence interval. This confirms that {\it Planck} data can indeed accommodate a low $S_8$ value within the DDM model at $1\sigma$.
Note also that the confidence intervals follow the Bayesian contours at $2\sigma$, indicating that volume effects are negligible at that level and that the 2$\sigma$ constraints are robust.

\section{\texttt{KiDS-1000} analysis}\label{sec:KiDS}
\begin{figure}
    \centering
    \includegraphics[scale=0.6]{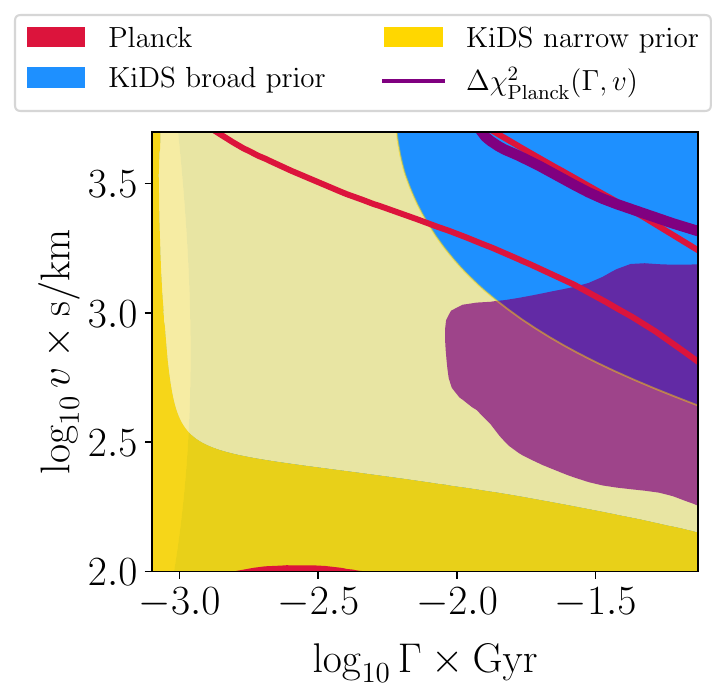}
    \caption{Bayesian constraints on the DDM model for \texttt{KiDS-1000} by changing the prior bounds on the primordial amplitude $A_s$ and the spectral index $n_s$. We also report the {\it Planck} Bayesian constraints and profile likelihood in red and purple as in \cref{fig:profile} for comparison. 
    }
    \label{fig:KiDS_nsprior}
\end{figure}

\subsection{What \texttt{KiDS-1000} Actually Measures}
\begin{figure}
    \centering
    \includegraphics[scale=0.65]{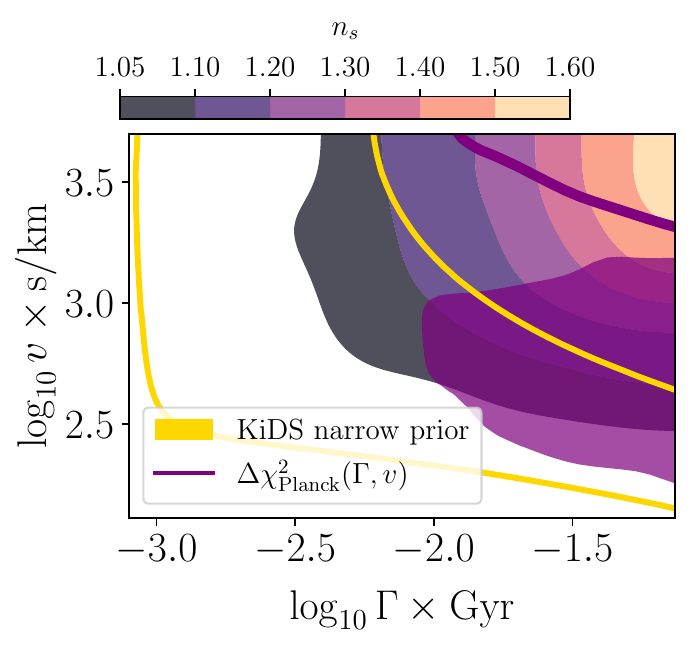}
    \caption{Smoothed 3D projection of the \texttt{KiDS-1000} posterior with broad priors on $A_s$ and $n_s$ in the $\Gamma$–$v$ plane, with the colour mapping representing the inferred scalar spectral index $n_s$. The yellow contour shows the posterior distribution from \texttt{KiDS-1000} with narrow priors for comparison. The purple contours show the results of the {\it Planck} profile likelihood. This figure illustrates the correlation between $n_s$ and the first diagonal in the plane $\Gamma$–$v$ that leads to the constraints presented in Ref.~\cite{Bucko:2023eix}.
    }
    \label{fig:KiDS_ns}
\end{figure}

To better understand the origin of the data constraining power and the $S_8$ tension, we show in \cref{fig:Pk_density} the constraints to the matter power spectrum obtained with the \texttt{KiDS-1000} weak lensing survey (left panels) and the {\it Planck} CMB survey (right panels). There, we have computed the power spectra of $500$ random points in the CDM (upper panels) and DDM (lower panels) models taken from their respective MCMC chains and plotted the density of points, all normalised to the best-fit $\Lambda$CDM model. 
We can, therefore, see the Bayesian constraints on the matter power spectrum for each scale directly by eye. 
Unfortunately, the {\it Planck}-DDM best-fit and part of the corresponding $1\sigma$ frequentist confidence region identified in \cref{sec:Planck_profile} lie outside the domain covered by the emulator $S^{\Gamma, v}$. In \cref{fig:KiDS_nsprior}, the overlap between the $1\sigma$ region and the trained domain of $S^{\Gamma, v}$ is shown as a purple contour. To visualise the variance of the matter power spectrum around the best-fit, we plot in red the density of $300$ power spectra evaluated within the accessible portion of the $1\sigma$ region.
Finally, the yellow band represents the main support of the $\sigma_8$ kernel $W(8~h/\mathrm{Mpc}, k) > 0.1$. The power spectra with less power in this region correspond to a lower $S_8$ value.

This figure is instructive in several ways: first, it shows that, for {\it Planck} and in the CDM case (top right panel), the power spectrum is highly constrained at all scales at a $\sim 1\%$ level. However, a large power suppression is allowed in the DDM model (bottom right panel), starting precisely around the $S_8$ kernel and extending to smaller scales. In that sense, the DDM model can alleviate the $S_8$ tension: it reduces $S_8$ without spoiling the fit to the CMB. However, one can see that the largest density of points reconstructed in a Bayesian analysis remains compatible with $\Lambda$CDM and no suppression. However, the best-fit to {\it Planck} data in the DDM model lies on the edge of the point density region, another illustration of the prior volume effect we studied.

Second, it is apparent that \texttt{KiDS-1000} data only constrain a range of small scales close to the $S_8$ band. However, the $S_8$ band does not perfectly overlap with the region of highest density of points and strongest constraints. In fact, it appears that \texttt{KiDS-1000} data would better constrain a kernel centred around $k\sim 0.5\,h/$Mpc. The data lacks effective constraining power at large scales, and the constraints are independent of the model (DDM or CDM). This explains why the reconstructed $S_8$ value in \cref{fig:bayesian} remains consistent across both models. A power suppression within the DDM model is degenerate with changes in $A_s$ and $n_s$; thus, opening up the DDM parameter space does not significantly affect the reconstructed matter power spectrum. Moreover, this suggests that constraints ``from \texttt{KiDS-1000} alone'' on the DDM model are prior dependent and that in the absence of any priors on $A_s$ and $n_s$, this data set does not significantly constrain the DDM model. This will be demonstrated explicitly in the next section. To break the degeneracy between all parameters in the DDM model, one needs to combine observations that probe all scales, \textit{e.g.}, combining {\it Planck} constraints to $A_s$ and $n_s$ with \texttt{KiDS-1000}.

\subsection{Prior Dependence of \texttt{KiDS-1000} Constraints}
\begin{figure}
    \centering
    \includegraphics[scale=0.6]{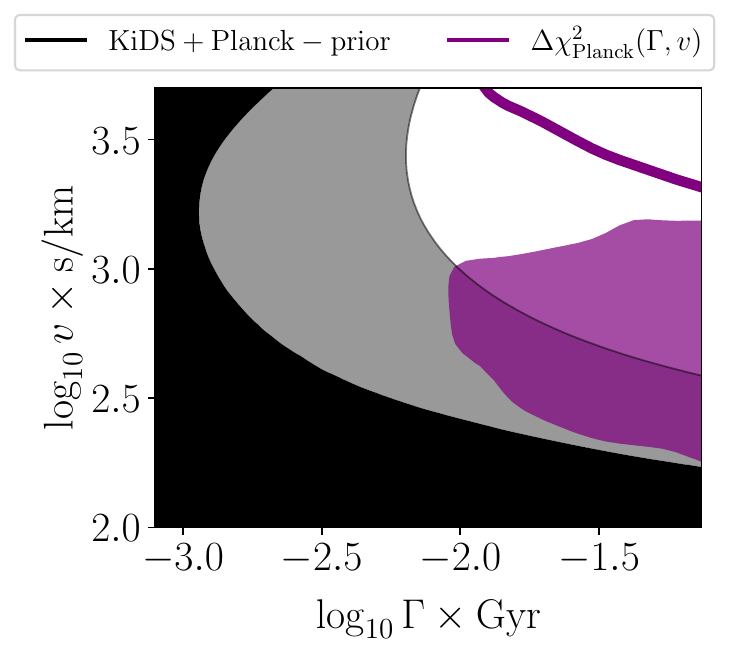}
    \caption{Constraints on the two additional parameters of the DDM model for {\it Planck}. The black contour shows the Bayesian \texttt{KiDS-1000} combined with {\it Planck} prior on $A_s$ and $n_s$. The purple contours show the {\it Planck} profile likelihood result. }
    \label{fig:final_constraints}
\end{figure}

\begin{figure}
    \centering
    \includegraphics[scale=0.4]{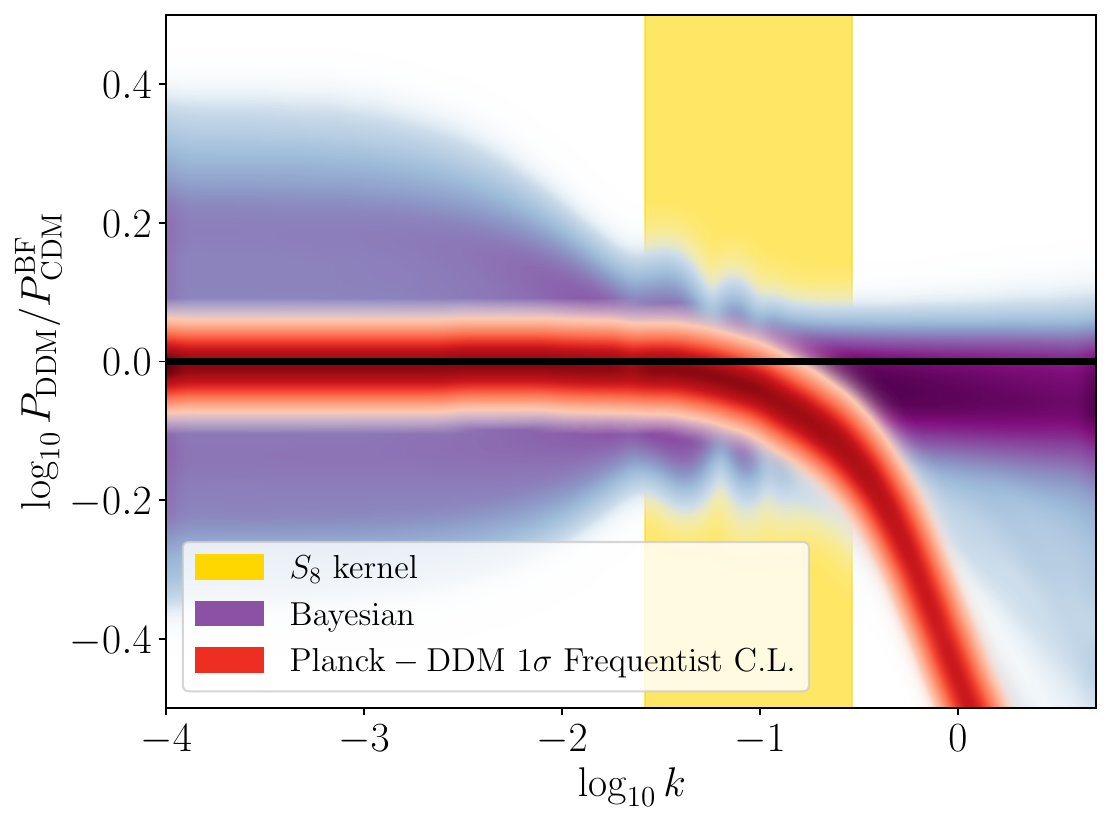}
        \caption{In blue, we show the Bayesian constraints on the matter power spectrum from \texttt{KiDS-1000}, combined with {\it Planck} priors on $A_s$ and $n_s$ in the DDM model. The red interval and yellow band are identical to those shown in \cref{fig:Pk_density}.}   
    \label{fig:final_PK_constraints}
\end{figure}

In \cref{fig:KiDS_nsprior} we show the reconstructed Bayesian posteriors from the analysis of \texttt{KiDS-1000} data alone in the $\{\log_{10}\Gamma,\log_{10} v\}$ plane. To illustrate the importance of the priors set on $A_s$ and $n_s$ for the constraints derived on the DDM parameters, we show results for two different sets of priors, given in \cref{tab:priors}.
\begin{table}[h]
    \centering
    \begin{tabular}{|c|c|c|}
    \hline
    Parameter & Broad prior & Narrow prior  \\
    \hline\hline
        $n_s$ & $U(0.5, 2)$ & $U(0.87, 1.07)$ \\
        $10^9 A_s$  & $U(0.1, 23.7)$ & $U(0.5, 5)$\\
    \hline
    \end{tabular}
    \caption{Uniform ($U(x_{\rm min}, x_{\rm max})$) priors on $A_s$ and $n_s$ set in the analysis of \texttt{KiDS-1000} data for the DDM model.}
    \label{tab:priors}
\end{table}
The narrow prior range is similar to that used in Ref.~\cite{Bucko:2023eix}. We further compare these constraints with those from {\it Planck} derived both in a Bayesian and a frequentist framework.
Note that the emulator $S$ of \cref{eq:pk_nl} does not span the whole range of $\Gamma$ and $v$ that we studied for {\it Planck}. In particular, the best-fit obtained with {\it Planck} data corresponds to a value of $\log_{10} \Gamma$ slightly above its upper limit in the range of the emulator. 

One can see that when we use broad priors on $n_s$ and $A_s$ (in blue), we obtain no constraints in the range of parameters spanned by the emulator, while when we use narrow priors (in yellow) similar to Ref.~\cite{Bucko:2023eix}, we exclude a large region of the parameter space corresponding to high values of $\Gamma$ and $v$. This behaviour can be understood by looking at the correlation between the spectral index and the plan $\Gamma-v$, depicted in \cref{fig:KiDS_ns}. The coloured bands correspond to isocontours of $n_s$ values. It is evident that the combination of $\Gamma$ and $v$ is strongly correlated with the spectral index $n_s$: models with high values of both $\Gamma$ and $v$ tend to produce values of $n_s \gtrsim 1.6$. 
This demonstrates that, as anticipated from our discussion of \cref{fig:Pk_density}, one can compensate for the effect of a power suppression in the DDM model by adjusting the primordial spectral tilt. 

However, it is clear that imposing {\it Planck} priors on $A_s$ and $n_s$ in the \texttt{KiDS-1000} analysis can prevent the strong degeneracy that we have uncovered from affecting the resulting constraints on the DDM parameters. 
We show in \cref{fig:final_constraints} the impact of imposing {\it Planck} constraints to $n_s$ and $A_s$ as Gaussian priors through the black contours, with values provided in \cref{tab:param_stat}. All the upper right part of the $\Gamma-v$ plan is now consistently excluded. Interestingly, one can identify a region of the parameter space at the intersection of the $1\sigma$ confidence level built from {\it Planck} data and the $1-2\sigma$ constraints from \texttt{KiDS-1000}. This is precisely the region that was found to alleviate the $S_8$ tension in past studies \cite{Abellan:2020pmw,Abellan:2021bpx,Simon:2022adh,Bucko:2023eix}. As \texttt{KiDS-1000} data have recently been updated in \texttt{KiDS-Legacy} \cite{Stolzner:2025htz}, whether the new data would firmly rule out this region remains to be confirmed. This figure also illustrates the power of multi-probe and multi-scale cosmological analyses: the constraining power of {\it Planck} at large scales allowed to obtain accurate $A_s$ and $n_s$ constraints, while the constraints from \texttt{KiDS-1000} at small-scales constrain $S_8$ and therefore the DDM parameters. 

Summarising, our results are nicely illustrated by the combined Bayesian constraints on the matter power spectrum shown from \texttt{KiDS-1000} with {\it Planck} priors on $A_s$ and $n_s$, shown in \cref{fig:final_PK_constraints}. One can see that with such priors, the $S_8$-kernel does match the best constrained area, although it extends to even smaller scales. In that case, the {\it Planck} best-fit DDM model appears to be excluded as it predicts a too strong suppression above  $k \gtrsim 1\,h$/Mpc. However, constraints to the matter power spectrum relax at scales $\log_{10}k \gtrsim 0.5\,h$/Mpc. 
This result underscores the constraining power of large-scale structure data beyond the sole \( S_8 \) statistic and highlights the importance of consistent, physically motivated priors when combining early- and late-time cosmological observations to constrain DM properties.

\begin{table*}
\begin{tabular}{|l|cc|ccc|}
\hline
Dark matter model & \multicolumn{2}{|c|}{CDM} & \multicolumn{3}{c|}{DDM} \\
\hline
Dataset & {\it Planck} & \texttt{KiDS-1000} & {\it Planck} & \texttt{KiDS-1000} & \texttt{KiDS-1000}+{\it Planck} Prior \\
\hline
$\Omega_{\rm m}$ & $0.3114\pm 0.0055$ & $0.233^{+0.031}_{-0.092}$ & $0.3110\pm 0.0056$ & $0.226^{+0.040}_{-0.076}$ & $0.287\pm 0.037$ \\
$S_8$ & $0.826\pm 0.010$ & $0.757^{+0.033}_{-0.018}$ & $0.816^{+0.022}_{-0.0066}$ & $0.755^{+0.037}_{-0.018}$ & $0.768^{+0.020}_{-0.018}$ \\
$n_s$ & $0.9666\pm 0.0037$ & $1.05^{+0.11}_{-0.14}$ & $0.9668\pm 0.0037$ & $1.05^{+0.10}_{-0.16}$ & $0.9669\pm 0.0038$ \\
$\ln 10^{10}A_s$ & $3.046\pm 0.014$ & $3.83^{+1.1}_{-0.81}$ & $3.045\pm 0.015$ & $3.81\pm 0.79$ & $3.045\pm 0.015$ \\
$f_{\rm DDM}$ & n/a & n/a & $<_{2\sigma} -0.16$ & $<_{2\sigma} -0.14$ & $<_{2\sigma} -1.32$\\
\hline
\end{tabular}
\caption{Constraints at a 68\% confidence level (and $95\%$ confidence level for $f_{\rm DDM}$) on the cosmological parameter of interest. For the DDM model parameters, we provide the combination of the two parameters that is better constrained by the data.}\label{tab:param_stat}
\end{table*}

\begin{figure}
    \centering
    \includegraphics[scale=0.75]{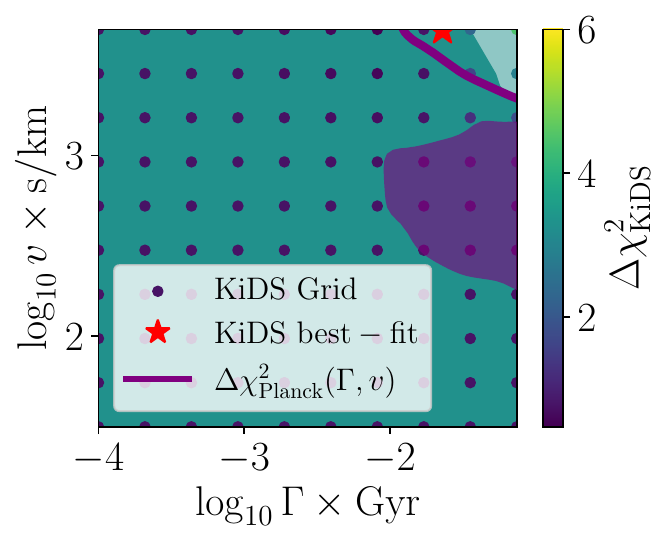}
    \caption{\texttt{KiDS-1000} 2D profile likelihood for $\Gamma-v$. The grid of DDM parameters used is shown as dots with colours corresponding to the minimum $\chi^2$ over all other parameters. The red star corresponds to the global minimum by also varying $\Gamma-v$ and the purple line shows again the {\it Planck} $1$ and $2\sigma$ frequentist confidence level.}
    \label{fig:KiDS_minimisation}
\end{figure}

\begin{figure*}
    \centering
    \includegraphics[scale=0.45]{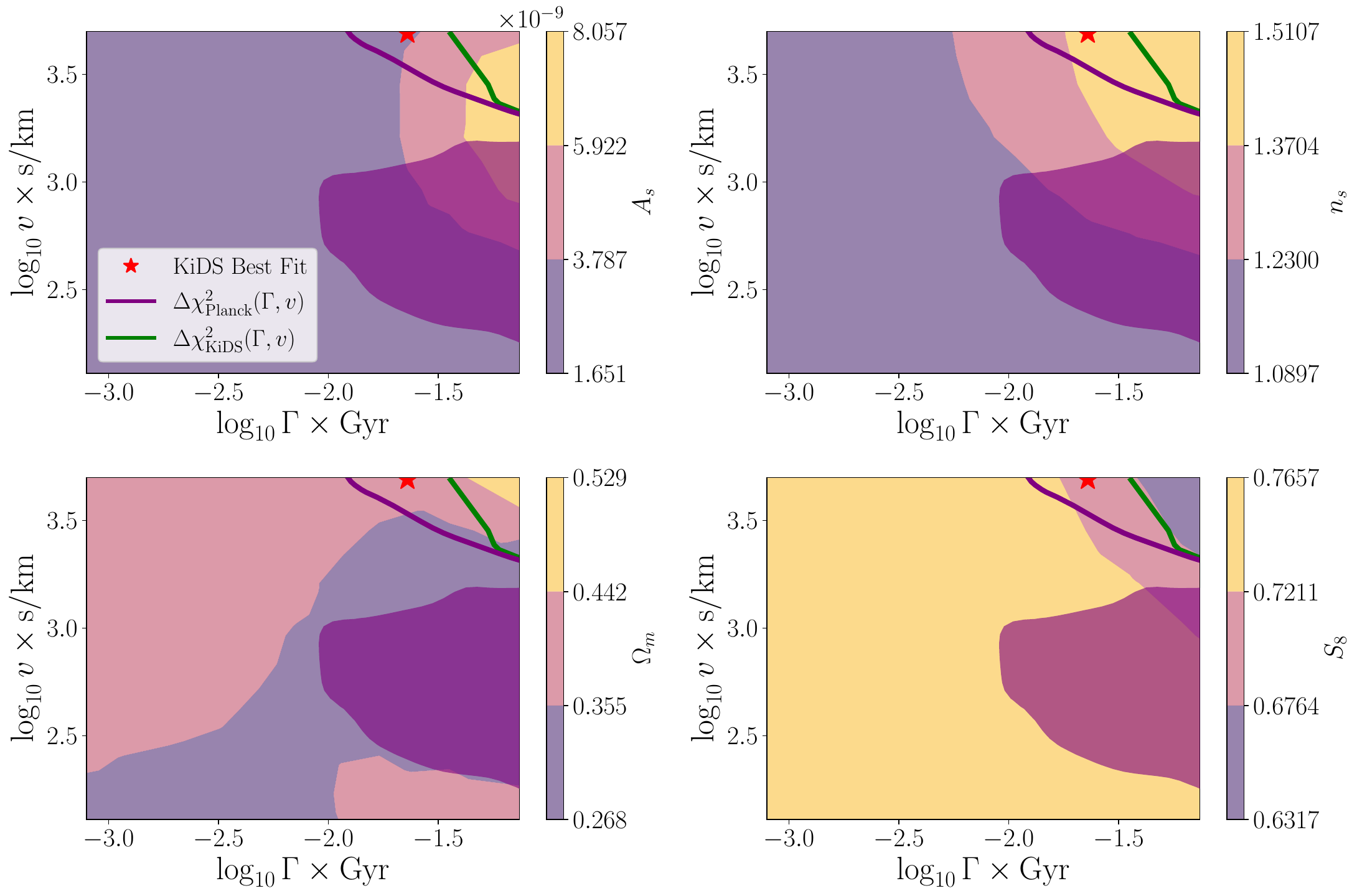}
    \caption{Isocontours of the best-fit values of $A_s$, $n_s$, $\Omega_{\rm m}$ and $S_8$ obtained when maximising the \texttt{KiDS-1000} likelihood on the 2D $\Gamma-v$ grid shown in \cref{fig:KiDS_minimisation}. }
    \label{fig:KiDS_minimisation_param}
\end{figure*}

\begin{figure}
    \centering
    \includegraphics[scale=0.7]{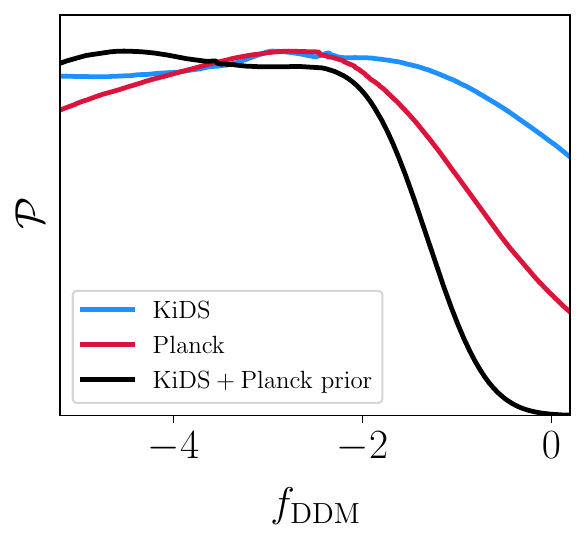}
    \caption{1D marginalised constraints on the combination of the two DDM parameters. The left limit of the plot corresponds to a CDM limit. We have improved the constraints by combining \texttt{KiDS-1000} (in blue) with {\it Planck} (in red) informed priors on $A_s$ and $n_s$. The $2\sigma$ upper limits are shown in \cref{tab:param_stat}}
    \label{fig:1D-final}
\end{figure}

\section{Conclusions} 
\label{sec:concl}
We have investigated the two-body Decaying Dark Matter (DDM) model as an extension of the standard cosmological scenario, in which cold dark matter particles decay into a warm daughter particle and a massless dark relativistic particle \cite{Wang:2012eka, Abellan:2021bpx}. This framework introduces two new physical parameters related to the decay dynamics: the decay rate $\Gamma$ and the momentum transfer $v$. For sufficiently small values of these parameters, the model reproduces standard $\Lambda$CDM predictions at early times, including CMB anisotropies with pure CDM component, while introducing distinctive late-time effects such as a time- and scale-dependent suppression of power on small scales. The ability of this model to relax late-time constraints on the properties of dark matter makes it a compelling candidate for addressing tensions in large-scale structure observations. Overall, it allows us to test the stability of dark matter on cosmological time scales, as unstable dark matter candidates naturally arise in many particle physics models \cite{Hambye:2010zb, Abazajian:2012ys, Drewes:2016upu, Berezinsky:1991sp, Covi:1999ty, Kim:2001sh, Feng:2003uy}. 

Using the matter power spectrum prediction in the linear regime, \cite{Blackadder:2014wpa, Abellan:2021bpx} and nonlinear regime calibrated on simulations \cite{Bucko:2023eix}, we have performed a standard Bayesian analysis of \textit{Planck}-2018 + BOSS-BAO + the Pantheon-Plus supernovae data (referred to as ``{\it Planck}'') on the one hand, and \texttt{KiDS-1000} data on the other hand, reproducing previous results that suggest a limited ability of the DDM model to alleviate the tension between {\it Planck} and \texttt{KiDS-1000} data.
However, we have pointed out that the lack of constraints on the new DDM parameters can easily induce large volume prior effects that could alter the consistency of the two datasets. 

To bypass this problem, we have performed for the first time a frequentist analysis of the DDM model in light of {\it Planck} data. Interestingly, we have found a 68\% confidence level that strongly differs from its Bayesian counterparts, with the hint of a slight preference for a specific region of the DDM parameter space that allows to reduce $S_8$. This confirms the presence of volume effects in the Bayesian analysis. In fact, we have found that the $S_8$ value reconstructed in the frequentist analysis is in agreement at $1\sigma$ with that of \texttt{KiDS-1000}, with a best-fit that is even lower, $S_8=0.70$. At the 95\% confidence level, frequentist and Bayesian constraints agree, and volume effects are negligible.  

In a second part, we have discussed the constraints set by the different surveys directly at the level of the matter power spectrum in order to better understand the origin of the $S_8$ tension and its resolution within the DDM model. We have shown that {\it Planck} data can constrain large scales precisely but allow for significant power suppression precisely in the range of scales probed by \texttt{KiDS-1000}. On the other hand, \texttt{KiDS-1000} do not allow to significantly constrain large-scales and small-scales independently (in particular $A_s$ and $n_s$). It only constrains a range of scales close to (but not exactly on) the $S_8$ kernel. As a result, large degeneracies with the DDM parameters prevent obtaining meaningful constraints on the DDM model in a \texttt{KiDS-1000} only analysis.

Concretely, we demonstrated that when using broad, uninformative priors, \texttt{KiDS-1000} data alone do not constrain the DDM parameters \((\Gamma, v)\), owing to a degeneracy between the spectral index \( n_s \) and \((\Gamma, v)\). This sensitivity to the prior volume reveals that previous \texttt{KiDS-1000} constraints on DDM, such as those from Ref.~\cite{Bucko:2023eix}, are significantly influenced by prior choices. We have thus performed a dedicated \texttt{KiDS-1000} analysis using \textit{Planck}-informed priors on \( n_s \) and \( A_s \). While this allows for a significant improvement of the constraints over the results from {\it Planck} and \texttt{KiDS-1000} alone, we find that a region of parameter space favoured in the {\it Planck} frequentist analysis at $68\%$ C.L. does survive the \texttt{KiDS-1000} constraints. This precisely matches the region that was found to alleviate the tension between both datasets in past studies \cite{Abellan:2020pmw,Abellan:2021bpx,Simon:2022adh,Bucko:2023eix}. 
It remains to be seen whether the updated \texttt{KiDS-Legacy} or \texttt{DES-Y3} data that show no tension with {\it Planck} would firmly exclude this region. 
Our work demonstrates the importance of combining probes that cover multiple scales to derive robust constraints on exotic dark matter properties. Looking forward, future weak lensing surveys such as \textit{Euclid} will dramatically improve sensitivity to the scale- and time-dependent suppression of structure induced by DDM models. Recent forecast analyses \cite{Euclid:2024pwi, FrancoAbellan:2024tbj} show that \textit{Euclid} will tighten constraints on the DDM parameter space by up to an order of magnitude, enabling a definitive test of the scenarios discussed in this work.

\section*{Acknowledgments} 
We thank Th\'eo Simon, Andreas Nygaard, Jozef Bucko, Aurel Schneider and Oliver Hahn for useful discussions. We are particularly thankful to Guillermo Franco Abell\'an for his thoughtful comments and suggestions on the draft.
EMT, TM, AP and VP are supported by funding from the European Research Council (ERC) under the European Union’s HORIZON-ERC-2022 (grant agreement no. 101076865). We gratefully acknowledge support from the CNRS/IN2P3 Computing Center (Lyon - France) for providing computing and data-processing resources needed for this work.

\appendix
\section{\texttt{KiDS-1000} Profile Likelihood}
\label{app:kids_pl}

We have performed the 2D profile likelihood analysis of the \texttt{KiDS-1000} data by following the same method as for \textit{Planck}, see \cref{eq:def_profile}. The result is shown in Fig.~\ref{fig:KiDS_minimisation}. The global best-fit obtained is indicated as a red star while the colours of the grid points axis correspond to $\Delta \chi^2_{\mathrm{KiDS}}(\Gamma, v)$. In dark and light green, we show respectively the $1$ and $2\sigma$ confidence level defined as $\Delta \chi^2_{\mathrm{KiDS}}(\Gamma, v) = 2.3$ and $\Delta \chi^2_{\mathrm{KiDS}}(\Gamma, v) = 6.17$. The \texttt{KiDS-1000} frequentist analysis seems to show that the Bayesian constraints obtained in Ref.~\cite{Bucko:2023eix} and reproduced with narrow priors on $n_s$ in \cref{fig:KiDS_nsprior} is a prior effect. 

However, we caution against strongly interpreting these results. Indeed, we find that the preferred value of $v$ tends to the highest value reachable by the emulator $v = 5000$~km$/$s. At the same time, the redshift dependence of the intrinsic alignment parameter called $\alpha_1$ in \texttt{CosmoSIS} and $\eta_{1}$ in \cite{Kilo-DegreeSurvey:2023gfr} tends to the highest allowed value by the default prior $5$. Unlike all the other nuisance parameters, $\alpha_1$ only has a flat prior, which appears to make it diverge during the minimisation procedure. Furthermore, the correlation between $\alpha_1$ and $n_s$ or $S_8$ is large, which means that $\alpha_1 \rightarrow 5$ can generate very low $S_8$ values ($<0.70$) or large $n_s$ ($>1.2$). In \cref{fig:KiDS_minimisation_param} we plot the isocontours for the best-fit parameters $A_s$, $n_s$, $\Omega_{\mathrm m}$ and $S_8$. For all these parameters, we observe that the upper right part gives high $A_s$, $n_s$ and $\Omega_{\rm m}$, that all together correspond to a very low $S_8$ value $(S_8<0.67)$. 

\section{Marginalised Statistics}
The constraints on DDM can be hard to represent with marginalised 1-dimensional statistics. Indeed, as long as CDM is not excluded, the two strongly correlated parameters $\Gamma$ and $v$ can take any value as long as the other one is small enough, \textit{i.e.} a large momentum transfer would have no effect if the particle lifetime is much larger than the age of the Universe, while a small lifetime would have no effect if the daughter particle remains cold. For this reason, the marginalised constraints on these parameters depend on the prior, which makes it difficult to interpret and compare between different analyses. As we have seen, \textit{e.g.} \cref{fig:profile,fig:final_constraints}, the data can provide constraints on the combination of $\Gamma$ and $v$ such that only simultaneously large values can be excluded. Hence, it makes sense to define the effective parameter 
\begin{equation}
    f_{\rm DDM} = \log_{10} \left(\frac{\Gamma}{\Gamma_0}\times \frac{v}{v_0}\right)\,,
\end{equation}
where $\Gamma_0$ and $v$ are arbitrary normalization values. Here, we choose the upper bounds of the emulator $\Gamma_0 = 1/13.5 $~Gyr$^{-1}$ and $v_0=5000$~km/s. Effectively, when $f_{\rm DDM} \ll 0$, we recover the CDM limit either by large lifetime or small momentum transfer. Constraints to this DDM parameter are shown in \cref{fig:1D-final}. One can see that the combination of \texttt{KiDS-1000} and {\it Planck} priors provide the strongest constraints to the model.

Finally, in Tab.~\ref{tab:param_stat}, we show the 1-dimensional marginalised constraints for all parameters, the two models we have considered, CDM and DDM, and the two datasets we have used {\it Planck} and \texttt{KiDS-1000}. The last column shows the combination of \texttt{KiDS-1000} and {\it Planck} informed priors on $A_s$ and $n_s$. 

\bibliography{main}
\end{document}